\documentclass{revtex4}
\usepackage{amssymb}
\usepackage{amsfonts}
\usepackage{amsmath}

\input{tcilatex}

\begin{document}

\title{Classical and quantum quasi-free position dependent mass; P\"{o}%
schl-Teller and ordering-ambiguity}
\author{S. Habib Mazharimousavi and Omar Mustafa}
\email{habib.mazhari@emu.edu.tr\\
omar.mustafa@emu.edu.tr}
\affiliation{Department of Physics, Eastern Mediterranean University, G. Magusa, north
Cyprus, Mersin 10 - Turkey,\\
Tel.: +90 392 6301078; fax: +90 3692 365 1604.}

\begin{abstract}
\textbf{Abstract: }We argue that the classical and quantum mechanical
correspondence may play a basic role in the fixation of the
ordering-ambiguity parameters. We use quasi-free position-dependent masses
in the classical and quantum frameworks. The effective P\"{o}schl-Teller
model is used as a manifested reference potential to elaborate on the
reliability of the ordering-ambiguity parameters available in the literature.

\end{abstract}

\maketitle

\section{Introduction}

The non-uniqueness in the presentation of the kinetic energy operator%
\begin{equation}
T=\frac{1}{4}\left[ m\left( x\right) ^{\boldsymbol{j}}pm\left( x\right)
^{k}pm\left( x\right) ^{l}+m\left( x\right) ^{l}pm\left( x\right)
^{k}pm\left( x\right) ^{j}\right] \text{; \ }j+k+l=-1,
\end{equation}%
of the von Roos position-dependent mass (PDM) Hamiltonian [1] has inspired
intensive research trends over the last few decades \cite%
{1,2,3,4,5,6,7,8,9,10,11,12,13,14,15,16,17,18,19,20,21,22,23,24,25,26,27,28,29,30,31,32,33,34,35,36,37,38}%
. Searching for some physically and/or mathematically acceptable
ordering-ambiguity parameters (i.e., $j$, $k$, and $l$), it is found that
the continuity conditions at the abrupt heterojunction between two crystals
suggest that $j=l$, otherwise the heterojunction behaves like impenetrable
barrier at which the wave functions vanish (cf., e.g., \cite{11,12,28,29,32}
for some details on this issue). In their study of the classical and quantum
mechanical correspondence on the PDM harmonic oscillator, Cruz et al. \cite%
{2} have shown that the special ordering $j=l$ $=-1/4$ and $k=-1/2$ in the
quantum picture is the one that gives rise to the potential term that is the
same as the classical PDM oscillator (with no ordering-ambiguity conflict).
Mustafa and Mazharimousavi \cite{12} have used a PDM pseudo-momentum
operator along with an intertwining process and have shown that such special
ordering (i. e., $j=l$ $=-1/4$ and $k=-1/2$) is a strictly determined
ordering. Similar arguments were reported by Cruz and Rosas-Ortiz in \cite%
{13}, by Koc et al. \cite{28} and by Bagchi et al. \cite{34}.

We contemplate, however, that the classical and quantum mechanical
correspondence may just be "the-other-way-around" in the ordering-ambiguity
parametric fixation process as to which ordering is to be classified as
"good" or "to-be-discarded". In this work, strictly speaking, we argue that
the fixation of the ordering-ambiguity parameters may, very well, be sought
through the classical observations of a given free PDM-particle moving under
the influence of its own internally byproducted force field (hence the
notion of quasi-free PDM-particle is unavoidably in point). That is, if the
Lagrangian descendent equations of motion for a classical PDM-particle
suggest that such classical particle is confined to move within a specific
range, then bound-states solution should quantum mechanically be feasibly
observed for the same PDM-particle. The opposite would also hold true, a
classical PDM-particle that is unconfined to move within any specific finite
range would correspond to a free particle textbook solution in quantum
mechanics.

Under such simplistic classical and quantum mechanical correspondence, we
organize our paper as follows. In section II, we give the essentials of a
classical "free" PDM-particle by choosing PDM functions at random for
illustration purposes. We discuss the byproducted forces introduced by the
PDM-particle itself. In the same section, we choose a 1D PDM-function where
the corresponding quantum mechanical effective potential is of a P\"{o}%
schl-Teller-type and reflect the classical and quantum mechanical
observations on the admissibility of the ordering-ambiguity parameters. In
section III, we discuss quasi-free PDM classical particle in 2D plane-polar
coordinates and use some power-law-type PDM-function as constructive
examples. Again, we use a 2D PDM-function that yields a P\"{o}%
schl-Teller-type effective potential in quantum mechanics. Classical and
quantum mechanical observations are reflected on the admissibility of the
ordering-ambiguity parameters. Our concluding remarks are given in section
IV.

\section{Quasi-free PDM classical particle in one-dimension}

In one-dimension (1D), the Lagrangian for a classical particle with
position-dependent mass (PDM), $m\left( x\right) $, moving in a free-force
field (i.e., $V\left( x\right) =0$) is given by 
\begin{equation}
\mathcal{L}\left( x,\dot{x}\right) =\frac{1}{2}m\left( x\right) \dot{x}^{2}.
\end{equation}%
Where only the kinetic energy term is involved in the Lagrangian. The
equation of motion associated with such a Lagrangian is%
\begin{equation}
m\left( x\right) \ddot{x}+\frac{1}{2}m^{^{\prime }}\left( x\right) \dot{x}%
^{2}=0,
\end{equation}%
where prime denotes derivative with respect to $x$ and dot denotes
derivative with respect to time $t$. Obviously, one observes that whilst the
PDM-particle is moving in an externally free-force field, its motion is
influenced by the effect of its own internally byproducted force field.
Unlike the classical free particle with constant mass and zero acceleration,
the quasi-free PDM-particle exhibits a deceleration or an acceleration of
the form%
\begin{equation}
\ddot{x}=-\frac{1}{2}\frac{m^{\prime }\left( x\right) }{m\left( x\right) }%
\dot{x}^{2},
\end{equation}%
depending on whether the byproducted force is damping or anti-damping,
respectively. The sign of the ratio $m^{^{\prime }}\left( x\right) /m\left(
x\right) $ determines the nature of the byproducted force. That is, when $%
m^{^{\prime }}\left( x\right) /m\left( x\right) >0$ the byproducted force is
a damping force (of frictional nature that slows down the particle) and when 
$m^{^{\prime }}\left( x\right) /m\left( x\right) <0$ the byproducted force
is anti-damping (speeds up the PDM-particle). Moreover, when $m^{^{\prime
}}\left( x\right) /m\left( x\right) =0$ the classical constant mass settings
are recovered. Such constant mass setting should not be considered here to
avoid triviality.

Nevertheless, in a straightforward manner one may show that Eq.(4) can be
rewritten as%
\begin{equation}
\frac{\ddot{x}}{\dot{x}^{2}}=-\frac{1}{2}\frac{m^{\prime }\left( x\right) }{%
m\left( x\right) }\Longrightarrow \frac{d\dot{x}}{\dot{x}}=-\frac{1}{2}\frac{%
dm\left( x\right) }{m\left( x\right) }.
\end{equation}%
Which upon integration would (with $x\left( t=0\right) =x_{0}$ and $\dot{x}%
\left( t=0\right) =\dot{x}_{0}$) result%
\begin{equation}
\int\limits_{\dot{x}_{0}}^{\dot{x}}\frac{d\dot{x}}{\dot{x}}=-\frac{1}{2}%
\int\limits_{x_{0}}^{x}\frac{dm\left( x\right) }{m\left( x\right) }%
\Longrightarrow \sqrt{m\left( x\right) }\dot{x}=\sqrt{m\left( x_{0}\right) }%
\dot{x}_{0}.
\end{equation}%
Using $m_{0}=m\left( x_{0}\right) $, $m\left( x\right) =m_{0}M\left(
x\right) $, $p_{x}=m\left( x\right) \dot{x}$ for the linear momentum, and $%
p_{x,0}=m_{0}\dot{x}_{0}$ for the initial linear momentum, we may then
recast (6) as%
\begin{equation}
p_{x}=p_{x,0}\sqrt{M\left( x\right) }.
\end{equation}%
Clearly, this result implies that the linear momentum of the PDM-particle is
not conserved (unlike the case where the linear momentum is conserved for a
free-particle with constant mass). Nevertheless, Eq. (6) suggests that only
the linear momentum of the square root of the PDM of a classical quasi-free
particle is conserved (in a metaphoric way so to speak). That is, if we
defined a new physical quantity $\Pi =\sqrt{m\left( x\right) }\dot{x}$ as
the PDM-quasi-linear momentum, then by the virtue of (6) one may conclude
that this new physical quantity $\Pi \equiv \Pi \left( x,\dot{x}\right) $ is
conserved, i.e.,%
\begin{equation}
\Pi \left( x,\dot{x}\right) =\Pi _{0}\left( x_{0},\dot{x}_{0}\right) .
\end{equation}%
However, a point canonical transformation (PCT) of the form%
\begin{equation}
q^{\prime }\left( x\right) =\sqrt{m\left( x\right) }
\end{equation}%
would imply that our Lagrangian in (2) transforms into%
\begin{equation}
\mathcal{L}\left( q,\dot{q}\right) =\frac{1}{2}\dot{q}^{2}.
\end{equation}%
Our new Lagrangian $\mathcal{L}\left( q,\dot{q}\right) $\ now represents the
Lagrangian of a constant "\textit{unit mass}" moving in $q$-space with a
constant (i.e., conserved) linear momentum $p_{q}=\dot{q}$ and subjected to
a force-free field $\ddot{q}=0$. In this case, one may (with $q_{0}=q\left(
x_{0}\right) =0$ for simplicity) write%
\begin{equation}
p_{q}=p_{q0}\Longrightarrow \dot{q}=\dot{q}_{0}\Longrightarrow q\left(
x\right) =\dot{q}_{0}\left( x_{0}\right) t=\Pi _{0}\left( x_{0},\dot{x}%
_{0}\right) t.
\end{equation}%
Then, the equation of motion in (3) along with the results (4)-(8) are
recovered. One should notice that (11) gives the trajectory $x$ of the PDM
in terms of the initial quasi-linear momentum $\Pi _{0}$ and the time $t$.

Therefore, the motion of a PDM-particle in an externally free-force field
(i. e., $V\left( x\right) =0$) may experience the effect of some damping or
anti-damping forces produced by the PDM-particle itself. Such byproducted
forces allow the PDM-particle to exhibit accelerated or decelerated motions
as shall be seen in the forthcoming illustrative examples.

\subsection{An exponential-type 1D classical-PDM}

Let us consider, for example, the set of 1D PDM-particles $m\left( x\right) $
that satisfies the condition%
\begin{equation}
\frac{m^{\prime }\left( x\right) }{m\left( x\right) }=2Ax^{n},
\end{equation}%
where $A$ and $n$ are two constants so that $%
\mathbb{R}
\ni A\neq 0$, and $0\leq n\in 
\mathbb{N}
$. Such setting would form a special set of PDM-particles with mass
functions given by%
\begin{equation}
m\left( x\right) =m_{0}\exp \left( \frac{2A}{n+1}x^{n+1}\right) .
\end{equation}%
The quasi-linear momentum conservation condition in (8), in turn, yields
(with $x_{0}=0$ for simplicity) a velocity of the form%
\begin{equation}
\dot{x}=\dot{x}_{0}\exp \left( -\frac{A}{n+1}x^{n+1}\right) .
\end{equation}%
The internally byproducted forces of the PDM-particles in (13) are therefore%
\begin{equation}
m\left( x\right) \ddot{x}=-A\,m\left( x\right) \,\dot{x}^{2}x^{n}=-A\left(
m_{0}\dot{x}_{0}^{2}\right) x^{n}.
\end{equation}%
Obviously, such forces represent a set of non-static (i.e., depends on the
initial velocity) position-dependent retarding or anti--retarding type
forces that slow down or speed up the PDM-particles of (13). For example,
the velocity decreases exponentially as $x\rightarrow \infty $ for $A>0$ and
increases exponentially as $x\rightarrow \infty $ for $A<0$ . Yet, when the
PDM-particle is moving in the negative $x$-direction (i.e., $\dot{x}_{0}<0$%
), then even and odd values of $n$ along with the value of $A$ in (14) would
determine the nature of the byproducted forces. Moreover, the PDM-particle
(13) should be given a non-zero initial velocity $\dot{x}_{0}$ in order to
move under the influence of its own byproducted force field.

For the sake of illustration, we now take $n=0$ in (13) to obtain%
\begin{equation}
m\left( x\right) =m_{0}\exp \left( 2Ax\right) ,
\end{equation}%
\begin{equation}
\dot{x}=\dot{x}_{0}\exp \left( -Ax\right) ,
\end{equation}%
and%
\begin{equation}
m\left( x\right) \ddot{x}=-A\left( m_{0}\dot{x}_{0}^{2}\right) .
\end{equation}%
Next, we consider $n=1$ in (13) to imply%
\begin{equation}
m\left( x\right) =m_{0}\exp \left( Ax^{2}\right) ,
\end{equation}%
\begin{equation}
\dot{x}=\dot{x}_{0}\exp \left( -\frac{A}{2}x^{2}\right) ,
\end{equation}%
and%
\begin{equation}
m\left( x\right) \ddot{x}=-A\left( m_{0}\dot{x}_{0}^{2}\right) x.
\end{equation}%
For both examples (i.e., even and odd $n$ values), one clearly observes that
for $\dot{x}_{0}>0$ ($\dot{x}_{0}<0$) and $A>0$ ($A<0$), the PDM-particle
(13) starts form $x_{0}=0$ and moves a finite distance in the positive
(negative) $x$-direction (but never reaches large enough distances).
Moreover, for $\dot{x}_{0}>0$ and $A<0$ the PDM-particles in (16) and (19)
start from $x_{0}=0$ but this time speed up exponentially to infinity (i.e., 
$\dot{x}\rightarrow +\infty $) in the positive $x$-direction. However,
whilst for $\dot{x}_{0}<0$ and $A>0$ the PDM-Particle in (16) starts from $%
x_{0}=0$ and\ speeds up exponentially to an infinite speed (i.e., $\dot{x}%
\rightarrow -\infty $) in the negative $x$-direction, the PDM-particle in
(19) would still move a finite distance in the negative $x$-direction. This
is documented in the manifestly retarding or anti--retarding byproducted
forces in (18) and (21). Similar scenario tendencies would repeat themselves
for even and odd values of $n$ in (12).

\subsection{An exactly solvable 1D-PDM; Classical and quantum mechanical
observations}

In this section, we consider the PDM-particle model%
\begin{equation}
m\left( x\right) =\frac{m_{0}}{\left( 1+B^{2}x^{2}\right) ^{2}},
\end{equation}%
and report some classical and quantum mechanical points of view. It should
be mentioned here that Mustafa and Mazharimousavi \cite{16} have provided a
quantum mechanical $d$-dimensional recipe and reported exact solution for a
specific ordering-ambiguity parametric set. In this work, however, we keep
the ambiguity-parameters as they are and discuss the validity for these
parameters.

\subsubsection{Classical mechanical observations}

Following our classical mechanical proposal above for the PDM-particle (22),
one would use the quasi-linear momentum conservation condition in (8) to
obtain%
\begin{equation}
\dot{x}=\dot{x}_{0}\left( 1+B^{2}x^{2}\right) ,
\end{equation}%
and then use (4) along with (23) to get%
\begin{equation}
\ddot{x}=2B^{2}\dot{x}_{0}^{2}\,\left[ x\left( 1+B^{2}x^{2}\right) \right] .
\end{equation}%
There should be no doubt that such a classical PDM-particle would have an
asymptotically infinite speed as $x\rightarrow \pm \infty $. A
straightforward integration of (23) would imply that 
\begin{equation}
x=\frac{1}{B}\tan \left( B\dot{x}_{0}t+\arctan \left( Bx_{0}\right) \right)
\in \left( -\infty ,\infty \right) .
\end{equation}%
The PDM byproducted-force is an anti-retarding/anti-damping force that
causes an acceleration that grows asymptotically to infinity with a growing $%
x$ (i.e., $\ddot{x}\rightarrow \pm \infty $ as $x\rightarrow \pm \infty $).
The PDM-particle (22) is not confined to move within a specific range,
therefore.

\subsubsection{Quantum mechanical observations}

Now we consider the PDM-particle of (22) using the quantum mechanical von
Roos Hamiltonian operator $H=T$ for a free-particle (i.e., $V\left( x\right)
=0$). Here, $T$ is given in (1), and $p=-i\partial _{x}$ (with $\hslash =1$%
). In a straightforward manner, one may show that the corresponding Schr\"{o}%
dinger equation%
\begin{equation}
H\psi \left( x\right) =E\psi \left( x\right) ,
\end{equation}%
would transform (using the substitution $\psi \left( x\right) =m\left(
x\right) ^{1/4}\varphi \left( q\right) $ along with the point canonical
transformation (9)) into%
\begin{equation}
\frac{1}{2}\left( -\partial _{q}^{2}+a\frac{m^{\prime \prime }\left(
x\right) }{m\left( x\right) ^{2}}-b\frac{m^{\prime }\left( x\right) ^{2}}{%
m\left( x\right) ^{3}}\right) \varphi \left( q\right) =E\varphi \left(
q\right) ,
\end{equation}%
with%
\begin{equation}
a=\frac{1}{4}\left( 1+2k\right) \text{, \ }b=\left[ \frac{9}{16}+j\left(
j+k+1\right) +k\right] ,
\end{equation}%
and $l$ is eliminated using the von Roos constraint in (1). This (together
with (22)) yields%
\begin{equation}
\frac{1}{2}\left( -\partial _{q}^{2}+\frac{4B^{2}}{m_{0}}\left( 5a-4b\right)
\tan ^{2}\left( \frac{Bq}{\sqrt{m_{0}}}\right) -\frac{4a}{m_{0}}B^{2}\right)
\varphi \left( q\right) =E\varphi \left( q\right) .
\end{equation}%
Now we introduce the change of variables of the form $z=Bq/\sqrt{m_{0}}$ to
obtain%
\begin{equation}
\left( -\frac{1}{2m_{0}}\partial _{z}^{2}+\frac{2\left( 5a-4b\right) }{%
m_{0}\cos ^{2}\left( z\right) }\right) \varphi \left( z\right) =\mathcal{E}%
\varphi \left( z\right) ,
\end{equation}%
where%
\begin{equation}
\mathcal{E=}\frac{1}{B^{2}}E+\frac{4}{m_{0}}\left( 3a-2b\right) .
\end{equation}

Obviously, this is the one-dimensional Schr\"{o}dinger equation for a P\"{o}%
schl-Teller type effective-potential%
\begin{equation}
V_{eff}\left( z\right) =\frac{2\left( 5a-4b\right) }{m_{0}\cos ^{2}\left(
z\right) }=\frac{1}{2m_{0}}\frac{\lambda \left( \lambda -1\right) }{\cos
^{2}\left( z\right) }
\end{equation}%
that admits exact bound state solution for $\lambda >1$ (i.e., $\left(
5a-4b\right) >0$). Such potential has impenetrable barriers manifested by
the singularities at $z=-\pi /2$ and $z=\pi /2$. Quantum mechanically
speaking, our PDM-particle in (22) would be confined to move between $x=-%
\sqrt{\left( e^{B\pi }-1\right) /\left( 2B^{2}\right) }$ and $x=\sqrt{\left(
e^{B\pi }-1\right) /\left( 2B^{2}\right) }$ for $\left( 5a-4b\right) >0$.

However, when $\left( 5a-4b\right) =0$ our PDM-particle (22) is set free and
the problem admits a textbook free particle solution. This would be in
agreement with the classical mechanical predictions mentioned above. Two
ordering-ambiguity parametric sets available in the literature satisfy this
case. They are, Zhu and Kroemer's \cite{35} ($j=l=-1/2$, $k=0$) and Mustafa
and Mazharimousavi's ($j=l=-1/4$, $k=-1/2$) \cite{12} orderings. On the
other hand, Ben Daniel and Duke's \cite{36} ($j=l=0$, $k=-1$) ordering
satisfies the bound-states condition and contradicts with the classical
mechanical observations, therefore. Yet, Gora and William's \cite{37} ($%
k=l=0 $, $j=-1$) and Li and Kuhn's \cite{38} ($k=l=-1/2$, $j=0$) orderings
yield $\left( 5a-4b\right) <0$ and would result in imaginary eigenstates
(hence, contradicting the classical observations)

\section{Quasi-free PDM classical particle in 2D; Plane-polar coordinates}

Using the plane-polar coordinates, the Lagrangian for a free classical
particle (i.e., moving in $V\left( r,\theta \right) =0$) with
position-dependent mass $m\left( r,\theta \right) =g\left( r\right) f\left(
\theta \right) $ reads%
\begin{equation}
\mathcal{L}=\frac{1}{2}m\left( r,\theta \right) \left( \dot{r}^{2}+r^{2}\dot{%
\theta}^{2}\right) =\frac{1}{2}g\left( r\right) f\left( \theta \right)
\left( \dot{r}^{2}+r^{2}\dot{\theta}^{2}\right) .
\end{equation}%
In this case, the descendent equations of motion are%
\begin{equation}
\frac{d}{dt}\left( \frac{\partial \mathcal{L}}{\partial \,\dot{r}}\right) =%
\frac{\partial \mathcal{L}}{\partial \,r}\Longrightarrow \,\ddot{r}+\left( 
\frac{g^{\prime }\left( r\right) }{g\left( r\right) }\right) \,\dot{r}%
^{2}+\left( \frac{f^{\prime }\left( \theta \right) }{f\left( \theta \right) }%
\right) \dot{r}\,\dot{\theta}=\frac{1}{2}\left( \frac{g^{\prime }\left(
r\right) }{g\left( r\right) }\right) \left( \dot{r}^{2}+r^{2}\dot{\theta}%
^{2}\right) +\,r\,\dot{\theta}^{2},
\end{equation}%
and%
\begin{equation}
\frac{d}{dt}\left( \frac{\partial \mathcal{L}}{\partial \,\dot{\theta}}%
\right) =\frac{\partial \mathcal{L}}{\partial \,\theta }\Longrightarrow 
\frac{d}{dt}\left[ g\left( r\right) f\left( \theta \right) \,r^{2}\,\dot{%
\theta}\right] =\frac{1}{2}g\left( r\right) f^{\prime }\left( \theta \right)
\left( \dot{r}^{2}+r^{2}\dot{\theta}^{2}\right) .
\end{equation}%
Whilst equation (34) for a constant $g\left( r\right) =1$, say, results in%
\begin{equation}
\,\ddot{r}+\left( \frac{f^{\prime }\left( \theta \right) }{f\left( \theta
\right) }\right) \dot{r}\,\dot{\theta}=r\,\dot{\theta}^{2}\Longrightarrow
f\left( \theta \right) \ddot{r}+f^{\prime }\left( \theta \right) \dot{r}\,%
\dot{\theta}=f\left( \theta \right) r\,\dot{\theta}^{2},
\end{equation}%
\newline
equation (35) for a constant $f\left( \theta \right) =1$, implies that%
\begin{equation}
\frac{d}{dt}\left[ g\left( r\right) \,r^{2}\,\dot{\theta}\right] =0.
\end{equation}

Let us consider, for example, a plane-polar free PDM-particle with only a
radially dependent mass (i.e., $m\left( r,\theta \right) =g\left( r\right) ,$
$f\left( \theta \right) =1$). In this case, equation (37) gives%
\begin{equation}
m\left( r\right) \,r^{2}\,\dot{\theta}=g\left( r\right) \,r^{2}\,\dot{\theta}%
=p_{\theta }=K\Longrightarrow g\left( r\right) \,r^{2}\,\dot{\theta}=g\left(
r_{0}\right) \,r_{0}^{2}\,\dot{\theta}_{0}
\end{equation}%
and equation (34) yields%
\begin{equation}
m\left( r\right) \ddot{r}=g\left( r\right) \ddot{r}\Longrightarrow \ddot{r}=-%
\frac{1}{2}\frac{g^{\prime }\left( r\right) }{g\left( r\right) }\dot{r}%
^{2}+\left( 1+\frac{g^{\prime }\left( r\right) }{2g\left( r\right) }r\right)
r\,\dot{\theta}^{2},
\end{equation}%
where $K$ is a constant. Using (38) and (39) along with the substitutions $%
\ddot{r}=\dot{r}\dot{r}^{\prime }$, $\dot{r}^{\prime }=d\dot{r}/dr$ and $%
u\left( r\right) =\dot{r}^{2}$ we obtain%
\begin{equation}
\left( g\left( r\right) u\left( r\right) \right) ^{\prime }=\frac{K^{2}}{%
r^{2}g\left( r\right) }\,\left[ \frac{2}{r}+\frac{g^{\prime }\left( r\right) 
}{g\left( r\right) }\right] \Longrightarrow g\left( r\right) u\left(
r\right) -g\left( r_{0}\right) u\left( r_{0}\right) =\int\limits_{r_{0}}^{r}%
\frac{K^{2}}{r^{2}g\left( r\right) }\,\left[ \frac{2}{r}+\frac{g^{\prime
}\left( r\right) }{g\left( r\right) }\right] dr.
\end{equation}%
Under the current polar PDM settings, the effect of the byproducted radial
force is obvious and documented in (39). Some aspects of such PDM methodical
proposal are illustrated through the following examples.

\subsection{A power-law type 2D classical-PDM}

Now we consider the set of PDM functions of the form%
\begin{equation}
m\left( r\right) =g\left( r\right) =m_{0}\left( \frac{r}{r_{0}}\right) ^{\nu
}=\Lambda r^{\nu },
\end{equation}%
where, $g\left( r_{0}\right) =m_{0}$, and $\Lambda =m_{0}r_{0}^{-\nu }$.
Equation (40) would then read%
\begin{equation}
g\left( r\right) u\left( r\right) -g\left( r_{0}\right) u\left( r_{0}\right)
=-\frac{K^{2}}{m_{0}r_{0}^{-\nu }}\left[ r^{-\nu -2}-r_{0}^{-\nu -2}\right] .
\end{equation}%
Consequently, one obtains%
\begin{equation}
m_{0}^{2}\dot{r}^{2}=\left( \frac{r_{0}}{r}\right) ^{\nu }\left[ m_{0}^{2}%
\dot{r}_{0}^{2}+\frac{K^{2}}{r_{0}^{2}}\right] -\frac{K^{2}}{r_{0}^{2}}%
\left( \frac{r_{0}}{r}\right) ^{2\nu +2}.
\end{equation}%
This result indicates that since the left-hand-side is zero or positive, so
should be the right-hand-side. That is,%
\begin{equation}
B_{0}^{2}\left( \frac{r_{0}}{r}\right) ^{\nu }-\tilde{K}^{2}\left( \frac{%
r_{0}}{r}\right) ^{2\nu +2}\geq 0\;;\;B_{0}^{2}=m_{0}^{2}\dot{r}_{0}^{2}+%
\tilde{K}^{2}\text{, and }\tilde{K}^{2}=\frac{K^{2}}{r_{0}^{2}}.
\end{equation}%
Which, in turn, implies that%
\begin{equation}
r^{\nu +2}\geq \frac{\tilde{K}^{2}r_{0}^{\nu +2}}{B_{0}^{2}}\Longrightarrow
\left\{ 
\begin{tabular}{l}
$r\geq r_{0}\left[ \frac{\tilde{K}^{2}}{B_{0}^{2}\,}\right] ^{1/\left( \nu
+2\right) }\;\text{; for }\nu >-2\smallskip $ \\ 
$r\leq r_{0}\left[ \frac{\tilde{K}^{2}}{B_{0}^{2}\,}\right] ^{1/\left( \nu
+2\right) }\;\text{; for }\nu <-2\smallskip $%
\end{tabular}%
\right. .
\end{equation}%
Clearly, one observes that for $\nu <-2$ our quasi-free radial PDM-particle
in (41) would be confined to move a maximum radial distance%
\begin{equation}
r_{\max }=r_{0}\left[ \frac{\tilde{K}^{2}}{B_{0}^{2}\,}\right] ^{1/\left(
\nu +2\right) },
\end{equation}%
whereas for $\nu >-2$ it would escape away to infinity.

For $\nu =-2$ the effect of $r\,\dot{\theta}^{2}$ in (39) would be
eliminated and equation (43) reads 
\begin{equation}
\dot{r}^{2}=\left( \frac{r_{0}}{r}\right) ^{-2}\dot{r}_{0}^{2}%
\Longrightarrow \dot{r}=\frac{r}{r_{0}}\dot{r}_{0},
\end{equation}%
and (38) yields%
\begin{equation}
\dot{\theta}=\frac{K}{m_{0}r_{0}^{2}}\Longrightarrow \dot{r}\frac{d\theta }{%
dr}=\frac{K}{m_{0}r_{0}^{2}}.
\end{equation}%
Next, substituting (47) in (48) implies that%
\begin{equation}
r=r_{0}\exp \left( \frac{m_{0}r_{0}\dot{r}_{0}}{K}\theta \right) .
\end{equation}%
This result suggests that when $\dot{r}_{0}$ is $\left( \pm \right) $ then $%
r $ grows up from $r_{0}$ to infinity as $\theta $ starts from zero to $\pm
\infty $, respectively, and a plane-spiral like trajectory is manifestly
introduced, therefore. When $\dot{r}_{0}$ is $\left( \pm \right) $ and $%
\theta $ starts from zero to $\mp \infty $, respectively, then $r$ shrinks
down from some $r_{0}$ to the center of the spiral-like trajectory.

\subsection{An exactly solvable 2D-PDM; Classical and quantum mechanical
observations}

In this section, we consider the radial PDM-particle model%
\begin{equation}
m\left( r,\theta \right) =g\left( r\right) =\frac{\tilde{C}}{\left(
1+C^{2}r^{2}\right) ^{2}}\,,
\end{equation}%
where $\tilde{C}=m_{0}\left( 1+C^{2}r_{0}^{2}\right) ^{2}$, and $g\left(
r_{0}\right) =m_{0}$. We keep the ordering-ambiguity parameters as they are
(i.e., we do not choose any specific ordering) and discuss again the
validity for these parameters.

\subsubsection{Classical mechanical observations}

Under such 2D plane-polar settings, our PDM-model in (50) would, when
substituted in (40), yield%
\begin{equation}
g\left( r\right) u\left( r\right) -g\left( r_{0}\right) u\left( r_{0}\right)
=-\frac{K^{2}}{\tilde{C}}\left[ \frac{\left( 1+C^{4}r^{4}\right) }{r^{2}}-%
\frac{\left( 1+C^{4}r_{0}^{4}\right) }{r_{0}^{2}}\right] .
\end{equation}%
Which would imply that%
\begin{equation}
g\left( r\right) \dot{r}^{2}=\tilde{a}^{2}-\frac{K^{2}}{g\left( r\right)
r^{2}},
\end{equation}%
where%
\begin{equation}
\tilde{a}^{2}=m_{0}\dot{r}_{0}^{2}+\frac{K^{2}}{g\left( r_{0}\right)
r_{0}^{2}},
\end{equation}%
is used for simplicity of calculations. One may then safely rewrite (with $%
\tilde{L}=\sqrt{\tilde{a}^{2}\tilde{C}/K^{2}}$) equation (52) as%
\begin{equation}
g^{2}\left( r\right) \dot{r}^{2}=\tilde{a}^{2}g\left( r\right) -\frac{K^{2}}{%
r^{2}}\geq 0\Longrightarrow C^{2}r^{2}-\tilde{L}\,r+1\leq 0.
\end{equation}%
This would, in turn, imply that 
\begin{equation}
\frac{\tilde{L}}{2C^{2}}-\frac{\tilde{L}}{2C^{2}}\sqrt{1-\frac{4C^{2}}{%
\tilde{L}^{2}}}\leq r\leq \frac{\tilde{L}}{2C^{2}}+\frac{\tilde{L}}{2C^{2}}%
\sqrt{1-\frac{4C^{2}}{\tilde{L}^{2}}}.
\end{equation}%
Consequently, such PDM-model (50) would be confined to move within a
specific range of $r$.

\subsubsection{Quantum mechanical observations}

For the PDM von Roos Hamiltonian in plane-polar coordinates for a
free-particle (i.e., $V\left( r,\theta \right) =0$), the Schr\"{o}dinger
equation with $\psi \left( r,\theta \right) =R\left( r\right) \Theta \left(
\theta \right) $ and $m\left( r,\theta \right) =g\left( r\right) $ separates
in a straightforward manner into a radial\ part and an angular part (cf.,
e.g., Mazharimousavi and Mustafa \cite{19} for more details on this issue).
The radial part of which is casted as%
\begin{equation}
\frac{R^{\prime \prime }\left( r\right) }{R\left( r\right) }+\left( \frac{1}{%
r\ }-\frac{g^{\prime }\left( r\right) }{g\left( r\right) }\right) \frac{%
R^{\prime }\left( r\right) }{R\left( r\right) }+\frac{\xi }{2}\left( \frac{%
g^{\prime }\left( r\right) }{g\left( r\right) }\right) ^{2}-\frac{\left(
k+1\right) }{2}\left[ \frac{g^{\prime }\left( r\right) }{rg\left( r\right) }+%
\frac{g^{\prime \prime }\left( r\right) }{g\left( r\right) }\right] -\frac{%
m^{2}}{r^{2}}=-2g\left( r\right) E,
\end{equation}%
where%
\begin{equation}
\xi =j\left( j-1\right) +l\left( l-1\right) -k\left( k+1\right) ,
\end{equation}%
and $m=0,\pm 1,\pm 2,\cdots $ is the magnetic quantum number. We now
substitute%
\begin{equation}
R\left( r\right) =r^{-1/2}g\left( r\right) ^{1/4}Q\left( q\left( r\right)
\right) ,
\end{equation}%
with a PCT $q^{\prime }\left( r\right) =\sqrt{g\left( r\right) }$ to obtain%
\begin{equation}
g\left( r\right) \frac{Q^{\prime \prime }\left( q\left( r\right) \right) }{%
Q\left( q\left( r\right) \right) }-\frac{\left( m^{2}-1/4\right) }{r^{2}}%
+\left( \frac{8\xi -7}{16}\right) \left( \frac{g^{\prime }\left( r\right) }{%
g\left( r\right) }\right) ^{2}-\frac{k}{2r}\left( \frac{g^{\prime }\left(
r\right) }{g\left( r\right) }\right) -\left( \frac{2k+1}{4}\right) \left( 
\frac{g^{\prime \prime }\left( r\right) }{g\left( r\right) }\right)
=-2g\left( r\right) E.
\end{equation}%
This equation, with $g\left( r\right) $ in (50),%
\begin{equation*}
q\left( r\right) =\left( \sqrt{\tilde{C}}/C\right) \arctan \left( Cr\right) 
\text{, and }z=Cq\left( r\right) /\sqrt{\tilde{C}}
\end{equation*}%
would read%
\begin{equation}
-Q^{\prime \prime }\left( z\right) +V_{eff}\left( z\right) Q\left( z\right)
=\eta Q\left( z\right) ,
\end{equation}%
where%
\begin{equation*}
\eta =\frac{2E\tilde{C}}{C^{2}}-8\xi +12k-1,
\end{equation*}%
and%
\begin{equation}
V_{eff}\left( z\right) =\frac{\left( m^{2}-1/4\right) }{\sin ^{2}z}-\left[ 
\frac{8\xi -8k-12-m^{2}+1/4}{\cos ^{2}z}\right] .
\end{equation}%
Obviously, Eq. (60) is the 1D Schr\"{o}dinger equation for a P\"{o}%
schl-Teller type effective potential (cf., e.g., [16]). Therefore, the
quantum particle in (50) moves between two infinite barriers at $z=0$ and $%
z=\pi /2$ if and only if the magnetic quantum number $m$ and the ordering
ambiguity parameters satisfy the conditions 
\begin{equation}
8\xi -8k-12-m^{2}+1/4<0\text{ , and }m^{2}-1/4>0\Longrightarrow \left\vert
m\right\vert =1,2,3,\cdots .
\end{equation}%
Although the S-states (i.e., states with the magnetic quantum number $m=0$)
are lost in the process of satisfying the second condition in (62), our
quantum particle is still confined to move between $r=0$ and $r=\sqrt{\left(
e^{C\pi }-1\right) /\left( 2C^{2}\right) }$. For $\left\vert m\right\vert
=1,2$, one observes that while conditions (62) are satisfied by Zhu and
Kroemer's \cite{35} ($j=l=-1/2$, $k=0$), Mustafa and Mazharimousavi's \cite%
{12} ($j=l=-1/4$, $k=-1/2$), Ben Daniel and Duke's \cite{36} ($j=l=0$, $k=-1$%
), and Li and Kuhn's \cite{38} ($k=l=-1/2$, $j=0$) orderings, Gora and
William's \cite{37} ($k=l=0$, $j=-1$) ordering fails to do so. However, for $%
\left\vert m\right\vert \geq 3$ we observe that all these orderings satisfy
conditions (62) and lead to bound-state solutions. Hence, quantum mechanical
observations would be in agreement with the classical mechanical
observations\ (55) when conditions (62) are satisfied.

\section{Concluding remarks}

In this work, the fixation of the ordering-ambiguity parameters of the von
Roos quantum mechanical kinetic energy operator (1) is sought through a
simplistic classical and quantum mechanical correspondence. The classical
observations of a given "quasi" free PDM-particle, moving under the
influence of its own internally byproducted force field, are used to reflect
on the corresponding quantum mechanical "quasi" free PDM-particle described
by the von Roos Hamiltonian. That is, if a classical PDM-particle is free to
move in an infinite range, then it should correspond to a free quantum
particle model and, therefore, admits free-particle wave solution. On the
other hand, if a classical particle is confined to move within a specific
finite range, then it should correspond to some bound-states problem in the
quantum mechanical picture (of course, with the proper boundary conditions
that may arise in the corresponding quantum mechanical treatment).

In the process, we have provided the essentials of classically "free"
PDM-particles in 1D and 2D along with some random illustrative examples. We
have discussed the by-producted forces introduced by such PDM-particles. We
have deliberately chosen the PDM-functions ((22) for the 1D and (50) for the
2D cases) that yielded P\"{o}schl-Teller-type effective potentials in the
quantum mechanical treatment ((32) and (61), respectively). In so doing, we
reflect our findings in the current work on Mustafa and Mazharimousavi's
results in \cite{16}. They have used similar PDM-particle subjected to
no-force field but rather trapped in its byproducted P\"{o}schl-Teller-type
effective potential and reported on the quantum bound states in
D-dimensions. Luckily, they have used Ben Daniel and Duke's \cite{36}
parametric set, $j=l=0$ and $k=-1$, which is the only parametric set that
leads to bound state solution as documented in the analysis of (32)
mentioned above. Strictly speaking, in both the 1D and 2D cases we have
found that Ben Daniel and Duke's ordering satisfies bound states conditions.
However, in the 1D case the bound-states quantum solution contradicts with
the classical observations and therefore finds no classical correspondence
(i.e., the classically PDM byproducted force is an
anti-retarding/anti-damping force and the particle is unconfined, $x\in
\left( -\infty ,+\infty \right) $).

Finally, nevertheless, we have observed that Zhu and Kroemer's \cite{35} ($%
j=l=-1/2$, $k=0$), and Mustafa and Mazharimousavi's \cite{12} ($j=l=-1/4$, $%
k=-1/2$) orderings have provided consistent quantum mechanical
correspondence to the classical observations for the 1D and 2D cases. This
would at least qualify these orderings as "reliable orderings". On the other
hand, however, the Gora and William's \cite{37} ($k=l=0$, $j=-1$) and Li and
Kuhn's \cite{38}\ ($k=l=-1/2$, $j=0$) orderings should be readily
disqualified. Not only on the grounds of the continuity conditions at the
abrupt heterojunction (where $j=l$ is sought) but also on the grounds of
failing, at least, to provide a consistent quantum correspondence to
classical observations in the 1D case. This does not necessarily mean to
disqualify Ben Daniel and Duke's ($j=l=0$, $k=-1$) as yet. More
investigations have to be made.


\begin{thebibliography}{99}
\bibitem{1} O. Von Roos, Phys. Rev. \textbf{B 27} (1983) 7547.

\bibitem{2} S. Cruz y Cruz, J Negro, L. M. Nieto, Phys. Lett. \textbf{A 369}
(2007) 400.

\bibitem{3} S. H. Mazharimousavi, J Phys \textbf{A}: Math. Theor.\textbf{41 (%
}2008\textbf{) }244016.

\bibitem{4} A. Schmidt, Phys. Lett. \textbf{A 353} (2006) 459.

\bibitem{5} A. Schmidt, J Phys \textbf{A}:\textbf{\ }Math. Theor. \textbf{42
(}2009\textbf{) }245304.

\bibitem{6} S. H. Dong, M. Lozada-Cassou, Phys. Lett. \textbf{A 337} (2005)
313.

\bibitem{7} I. O. Vakarchuk, J. Phys. \textbf{A}: Math. Gen. \textbf{38}
(2005) 4727.

\bibitem{8} C. Y. Cai, Z. Z. Ren, G. X. Ju, Commun. Theor. Phys. \textbf{43}
(2005) 1019.

\bibitem{9} B. Roy, P. Roy, Phys. Lett. \textbf{A 340} (2005) 70.

\bibitem{10} B. Gonul, M. Kocak, Chin. Phys. Lett. \textbf{20} (2005) 2742.

\bibitem{11} A. de Souza Dutra, C A S Almeida, Phys Lett. \textbf{A 275}
(2000) 25.

\bibitem{12} O. Mustafa, S. H. Mazharimousavi, Int. J. Theor. Phys \ \textbf{%
46} (2007) 1786.

\bibitem{13} S. Cruz y Cruz, O Rosas-Ortiz, J Phys \textbf{A}: Math. Theor. 
\textbf{42} (2009) 185205.

\bibitem{14} J. Lekner, Am. J. Phys. \textbf{75} (2007) 1151.

\bibitem{15} L. Jiang, L. Z. Yi, C. S. Jia, Phys. Lett. \textbf{A 345}
(2005) 279.

\bibitem{16} O. Mustafa, S. H. Mazharimousavi, Phys. Lett. \textbf{A 358}
(2006) 259.

\bibitem{17} A. D. Alhaidari, Phys. Rev. \textbf{A 66} (2002) 042116.

\bibitem{18} O. Mustafa, S. H. Mazharimousavi, J. Phys. \textbf{A}: Math.
Gen. \textbf{39} (2006) 10537.

\bibitem{19} S. H. Mazharimousavi, O. Mustafa, SIGMA \textbf{6 }(2010) 088.

\bibitem{20} B. Bagchi, A. Banerjee, C. Quesne, V. M. Tkachuk, J. \ Phys. 
\textbf{A}: Math. Gen. \textbf{38} (2005) 2929.

\bibitem{21} J. Yu, S. H. Dong, Phys. Lett. \textbf{A 325} (2004) 194.

\bibitem{22} C. Quesne, Ann. Phys. \textbf{321} (2006) 1221.

\bibitem{23} T. Tanaka, J. Phys. \textbf{A}: Math. Gen. \textbf{39} (2006)
219.

\bibitem{24} A. de Souza Dutra, J. Phys. \textbf{A}: Math. Gen. \textbf{39}
(2006) 203.

\bibitem{25} O. Mustafa, S. H. Mazharimousavi, Czech. J. Phys \textbf{56}
(2006) 967.

\bibitem{26} O. Mustafa, S. H. Mazharimousavi, Phys. Lett. \textbf{A 357}
(2006) 295.

\bibitem{27} O. Mustafa, S. H. Mazharimousavi, J Phys \textbf{A}: Math.
Theor. \textbf{41 (}2008\textbf{) }244020.

\bibitem{28} R. Koc, G. Sahinoglu, M. Koca, Eur. Phys. J. \textbf{B 48 }%
(2005) 583.

\bibitem{29} O. Mustafa, S. H. Mazharimousavi, Phys. Lett. \textbf{A 373}
(2009) 325.

\bibitem{30} O. Mustafa, S. H. Mazharimousavi Phys. Scr.\textbf{\ 82 }(2010)
065013.

\bibitem{31} O. Mustafa, J Phys \textbf{A}:\textbf{\ }Math. Theor. \textbf{%
43 (}2010\textbf{) }385310.

\bibitem{32} O. Mustafa, J Phys \textbf{A}: Math. Theor. \textbf{44 (}2011%
\textbf{) }355303.

\bibitem{33} Y. Hamdouni, J Phys \textbf{A}: Math. Theor. \textbf{44 (}2011%
\textbf{) }385301.

\bibitem{34} B. Bagchi, P. Gorain, C. Quesne and R. Roychoudhury, Mod. Phys.
Lett. \textbf{A 19} (2004) 2765.

\bibitem{35} Q. G. Zhu, H. Kroemer, Phys. Rev. \textbf{B 27} (1983) 3519.

\bibitem{36} D. J. Ben Daniel, C. B. Duke, Phys. Rev. \textbf{152} (1966)
683.

\bibitem{37} T. Gora, F. Williams, Phys. Rev. \textbf{177} (1969) 1179.

\bibitem{38} T. Li, K. J. Kuhn, Phys. Rev. \textbf{B 47} (1993) 12760.
\end{thebibliography}
\end{document}